
\NPrefs
\def\define#1#2\par{\def#1{\Ref#1{#2}\edef#1{\noexpand\refmark{#1}}}}
\def\con#1#2\noc{\let\?=\Ref\let\<=\refmark\let\Ref=\REFS
         \let\refmark=\undefined#1\let\Ref=\REFSCON#2
         \let\Ref=\?\let\refmark=\<\refsend}

\define\SEN
A. Sen, Phys. Rev. {\bf D32} (1985) 2102, Phys. Rev. Let. 55 (1985) 1846;
C. G. Callan, D. Friedan, E. Martinec and M. Perry, Nucl. Phys. {\bf B262}
(1985) 593.

\define\MS
S. Mukherji and A. Sen, Nucl. Phys. {\bf B363} (1991) 639.

\define\ASF
A. Sen, Phys. Lett. {\bf B252} (1990) 566.

\define\MY
R. C. Myers, Phys. Lett. {\bf B199} (1988) 371; I. Antoniadis, C. Bachas,
J. Ellis and D. Nanopoulos, Nucl. Phys. {\bf B328} (1989) 117; S. P. de
Alwis, J. Polchinski and R. Schimmrgk, Phys. Lett. {\bf 218} (1989) 449.

\define\CL
J. Cardy and C. Ludwig, Nucl. Phys. {\bf B285} (1987) 687.

\define\Z
A. B. Zamolodchikov, JETP Lett. {\bf 43} (1986) 731; Sov. J. Nucl. Phys.
{\bf 46} (1987) 1090.

\define\DDW
S. Das, A. Dhar and S. Wadia, Mod. Phys. Lett. {\bf A5} (1990) 799.

{}~\hfill\vbox{\hbox{TIFR/TH/92-11}\hbox{February, 1992}}\break

\def\cph{C_{\phi\phi\phi}}
\def\p{\partial_\sigma}
\def\lm{\lambda}
\def\si{\Sigma}

\title{ ON THE INTERPOLATING SOLUTIONS OF
STRING IN DIFFERENT BACKGROUNDS }

\author{ Sudipta Mukherji}
\address{ Tata Institute of Fundamental Research\break Homi Bhabha Road,
Bombay 400 005, India}

\abstract
We analyze the beta-function equations for string theory
in the case when the target space has
one spacelike (or timelike) direction and rest is some
conformal field theory (CFT) with
appropriate central charge and has one nearly marginal operator.
We show there always exists a space (time)
dependent solution which interpolates
between the original background and the background where CFT is replaced
by a new conformal field theory, obtained by perturbing CFT by the nearly
marginal operator.
\endpage

In this letter we investigate, in detail, the beta-function equations for
string in a background, which is a direct sum of two conformal field
theories ($CFT$'s). One of which ($CFT_1$) contains a nearly
marginal praimary operator $\phi$
of dimension $(1-h,1-h;h<<1)$ with corresponding
coupling $\lm$ and has central
charge $c_1 = 25 -3Q^2$. The other ($CFT_2$) is a conformal field theory
of spacelike scalar\foot{Our entire discussion will go through even when
we take a time like scalar field and the other, a CFT with central charge
$25+3Q^2$ as a background.}
 field $\si$
 with a background charge $Q$ at infinity such that stress energy tensor
$$
T_2(z) = - {1\over 2} (\partial\si\partial\si + Q\partial^2\si)
\eqn\tone
$$
has a central charge $c_2 = 1 + 3Q^2$.

With this background, the two dimensional world sheet action has the
form
$$
S_1 -\int d^2 z \lm(\sigma)\phi(z,\bar z) + S_{CFT_2}
\eqn\tfifteen
$$
where $\lm$ is the coupling constant associated with the nearly marginal
operator $\phi$,
$S_1$  is the $\lm$ independent part of the $CFT_1$ action and
$S_{CFT_2}$ is the corresponding
action for $CFT_2$. Here we have used $\sigma$ as the zero mode of $\si$.
The first two terms
in \tfifteen~defines a conformally invariant action when it is at the
fixed points of the renormalization group flow of the two dimensional
theory. These fixed points correspond to particular values of the coupling
parameter $\lm$. It is known \Z\CL~that the two nearby fixed points of this
action are at $\lm=0$ and $\lm = 2h/\cph + {\cal O}(h^2)$. The point
$(\lm =2h/\cph, \sigma = \infty)$
corresponds to a new conformal field theory
$CFT_1^\prime$ with different central charge
($c_1^\prime$) from the original with $\phi$ being an irrelavent operetor
of $CFT_1^\prime$. The change in the central charge can be calculated by
several means \CL\MS. That can be given by
$$
c_1 - c_1^\prime = {8h^3\over \cph^2} + {\cal O}(h^4).
\eqn\tsixteen
$$
Thus if $CFT_2^\prime$ denotes a new conformal field theory with stress
energy tensor
$$
T_2^\prime = -{1\over 2}(\p\si\p\si + Q^\prime \p^2\si).
\eqn\tseventeen
$$
Then $CFT_1^\prime \oplus CFT_2^\prime$ will describe a new background for
string theory provided the total central charge adds up to 26, i.e,
$$
1 + 3Q^\prime + 25 - 3Q^2 - {8h^3\over \cph^2} = 26
\eqn\teighteen
$$
or equivalently
$$
Q^\prime = (Q^2 + {8h^3\over 3 \cph^2})^{1\over 2} = Q + {8h^3\over 6
Q\cph^2} + {\cal O}(h^4).
\eqn\tnineteen
$$

In this letter we will show that there that
there exists a $\sigma$ dependent solution of the coupled set of beta-function
equations involving dilaton, graviton and $\lm$ which interpolates between
these two backgrounds.

Notice that if we iterprete $\si$ to be the Liouville coordinate, this
result holds equally well for minimal model coupled to gravity theories.

Given the system we can write down the $\sigma$-model action as,
$$
S ={1\over{8\pi\alpha}}\int d^2\zeta [g^{ab} G_{\sigma\sigma}
\partial_a\si\partial_b\si + \alpha R^{(2)}\Phi
 -\lambda\phi(\zeta)] + S_1.
\eqn\ttwo
$$
Here $G_{\sigma\sigma}$, $\Phi$ are the one component graviton and
dilaton respectively and
$S_1$ has been defined earlier. Equations of motion are obtained by setting
beta-functions to zero~\SEN. Those are:
$$
\p^2D -{1\over 2}G^{-1}\p G\p D - {1\over 2} Q G^{-1}\p G -(\p \lambda)^2
= 0
\eqn\tthree
$$
$$
2G^{-1}\p^2 D + G^{-1}(\p D + Q)^2 - G^{-1}\p G\p (D+Q\sigma) - Q^2 -
G^{-1}(\p\lm )^2 - 2h\lm^2 + {2\over 3}\cph\lm^3 = 0
\eqn\tfour
$$
and
$$
G^{-1}\p^2\lm + G^{-1}\p\lm\p (D +Q\sigma) -2h\lm + \cph\lm^2 -{1\over
2}G^{-2}\p G\p\lm = 0.
\eqn\tfive
$$
Here $\cph$ is the structure constant of $\phi$ in the operator product
expansion of $\phi$ with $\phi$ and $D$ is the shifted dilaton coupling
defined as $ D = \Phi - Q\sigma$.
In the above equations we have used $G_{\sigma\sigma} = G$ and
$G^{\sigma\sigma} = G^{-1}$.

As usual, these equations can be obtained from an effective field theory
action of the form:
$$
S_{eff} = \int e^{D+Q\sigma} {\sqrt G}[ - G^{-1}(\p D)^2 - 2G^{-1}Q\p D -
G^{-1}Q^2 -G^{-1}(\p\lm)^2 - 2h\lm^2 +{2\over 3}\cph \lm^3 - Q^2]
\eqn\tsix
$$
Varying this action with respect to corresponding fields $G$, $D$, $\lm$
we get,
$$
G^{-1}(\p D)^2 + 2QG^{-1}\p D + G^{-1}Q^2 + G^{-1}(\p\lm)^2 + {2\over
3}\cph \lm^3 - 2h\lm^2 - Q^2 = 0
\eqn\tseven
$$
$$\eqalign{
&2G^{-1} \p^2 D + G^{-1}(\p D)^2 + 2QG^{-1}\p D - G^{-1}(\p \lm)^2 -
2h\lm^2 + {2\over 3}\cph\lm^3 - Q^2\cr
& + G^{-1}Q^2 - G^{-2}\p G\p D - Q
G^{-2}\p G = 0\cr}
\eqn\teight
$$
and
$$
G^{-1}\p^2\lm + (Q+\p D)G^{-1}\p\lm - 2h\lm + \cph\lm^2 -{1\over 2}
G^{-1}\p G\p \lm = 0
\eqn\tnine
$$
respectively. Appropriate linear combinations of these equations reproduce
eqs.~\tthree~-~\tfive.
$S_{eff}$ being the low energy effective action
for string theory, it has to have general coordinate invariance. Under
general coordinate transformation dilaton and graviton transforms as
$$
D^\prime(\sigma^\prime) + Q\sigma^\prime = D(\sigma) + Q\sigma;~~
G^\prime(\sigma^\prime) = ({d\sigma^\prime\over d\sigma})^{-2} G(\sigma).
\eqn\tten
$$
So if we choose the transformed coordinate $\sigma^\prime = \Lambda(\sigma)$
where $\Lambda =\int {\sqrt G} d\sigma$, we see from \tten~that
$$
G^\prime(\sigma^\prime) = 1;~~
D^\prime(\sigma^\prime) = D(\sigma) + Q(\sigma-\int {\sqrt G}d\sigma).
\eqn\televen
$$
Now eqs. \tthree~\tfive~and \tseven~ in this particular gauge reduce to
simpler forms
$$
\p^2 D = (\p \lm)^2
\eqn\ttwelve
$$
$$
\p^2\lm +(Q+\p D)\p\lm - 2h\lm + \cph\lm^2 = 0
\eqn\tthirteen
$$
and
$$
(\p D)^2 + 2Q\p D + (\p \lm)^2 - 2h\lm^2 +{2\over 3} \cph\lm^3 = 0.
\eqn\tfourteen
$$
Here we have removed all the primes from $D$ and $\sigma$ for convenience.
Eq.\tthirteen~
can now be interpreted as a particle moving in a potential $-h\lm^2 +(1/3)
\cph\lm^3$ with a force
$-(Q+\p D)\p\lm$. The potential has a maximum at
$\lm = 0$ and minimum at $\lm = (2h/\cph)$. The question we would like to
ask is whether there exists at least one solution which interpolates
between these two points. If the answer to this question is yes then it
will tell us that the particle originally at coordinte $(\lm = 0,\sigma =
-\infty)$ will roll down to $(\lm = 2h/\cph, \sigma = \infty)$.\foot{If we
want the solution to lowest order in $h$ we can neglect the dilaton
derivative term in \tthirteen~and hence it becomes an independent
equation. Analysis for such case has been provided in \ASF\DDW.}

Now we will show there exists a solution of the coupled set of
equations \ttwelve~-~\tfourteen~with required buondary conditions as
mentioned above.

Near $\sigma=-\infty, \lm \rightarrow 0$
we can solve eqs. \ttwelve~and \tthirteen~
iteratively. We start with assuming $\p D$ is zero and linearize
\tthirteen~ near $\lm = 0$. That gives
$$
\p^2\lm +Q\p\lm - 2h\lm = 0
\eqn\ttwentyone
$$
with solution,
$$
\lm = e^{{2h\over Q}\sigma}.
\eqn\ttwentytwo
$$
Plugging this back in \ttwelve~ and using the boundary condition $\p D =
0$ at $\sigma = -\infty$, we get,
$$
\p D = h e^{{4h\over Q}\sigma}.
\eqn\ttwentythree
$$
Obviously the solution is consistent upto this order because when we use
\ttwentythree~in \tthirteen~we get correction in $\lm$ which is higher
power in $e^{2h\sigma /Q}$.

It is clear from \ttwentythree~that since $h\ge 0$, $\p D$ is positive
near $\sigma = -\infty$. Moreover using \ttwelve, we see that
it always increases with $\sigma$. Hence the term $(Q+\p D)$
in \tthirteen~is always positive. So \tthirteen~can be thought of as a
particle moving in a potential $-h\lm^2 +(1/3)\cph\lm^3$ with a large
damping force $(Q+\p D)\p \lm$. This certainly shows that the particle
will start from $\lm = 0$ and ultimately will settle down at $\lm =
2h/\cph$. This proves that there always exists one solution for $\lm$
which goes from one point to the other.

{}From \tfourteen~we can put a bound on $\p D$.
$$\eqalign{
\p D &= -Q + {\sqrt{ Q^2 - 2\{{1\over 2}(\p \lambda)^2 - h\lambda^2 +
{1\over 3}\cph\lambda^3\}}}\cr
&= -Q + {\sqrt {Q^2-2T-2V}}\cr}
\eqn\ttwentyfour
$$
where $T= {1\over 2} (\p\lambda)^2$, $V = -h\lambda^2 + {1\over 3}
\cph\lambda^3$. $V$ has the maximum value of $-4h^3/3\cph^2$ and since $T$ is
always positive,
$$
\p D\le {8h^3\over 6\cph^2Q}.
\eqn\ttwentyfive
$$
Here the equality holds only at the minimum of the potential.

As before we can analyze the nature of the solution at $\sigma = \infty$.
We can define shifted variables as
$$
\bar\lm = {2h\over \cph} - \lambda
\eqn\ttwentysix
$$
and
$$
{\overline{\p D}} = {8h^3\over 6\cph^2 Q}-\p D.
\eqn\ttwentyseven
$$
Substituating \ttwentysix~in \tthirteen, we get,
$$
\p^2\bar\lm + (Q+{8h^3\over 6\cph^2 Q})\p\bar\lm + 2h\bar\lm = 0.
\eqn\ttwentyeight
$$
(Here we have linearized the solution and neglected the term proportional
to ${\overline{\p D}}{\p{\bar\lm}}$.) Equation~\ttwentyeight~can now be
interpreted as a particle moving in presence of new background charge
$Q^\prime = Q + 8h^3/6\cph^2 Q$.
The new background can of course be gotten by giving a
$\sigma$-dependent expectation value to dilaton
field of the form ( see also \MY\MS ) :
$$
D = (Q^\prime - Q)\sigma = {8h^3\over 6\cph^2 Q}\sigma
\eqn\ttwentynine
$$
since \ttwentynine~is always a solution of \ttwentyfive~at the minimum of
the potential. This is precisely the change in the background charge we
expect as has been discussed at the begining ( see eq.~\tnineteen).

All along our discussion we have assumed $Q$ to be positve. Hence the
particle dynamics is governed by a large damping force in a potential
discussed before. This is the scenario for string in an expanding universe.
On the other hand, if $Q$ is negative the damping force changes the sign
keeping the potential same. This corresponds to
string in a contracting universe. In this case the solution will
interpolate between $CFT_1^\prime \oplus CFT_2^\prime$ at $\sigma=-\infty$
to $CFT_1 \oplus CFT_2$ at $\sigma=\infty$.

\noindent{\bf{Acknowledgements}}:
I am delighted to thank A. Sen for several useful discussions. His insight
and constant interest during this work
were of great importance to me. It is also a
pleasure to thank S. Das, S. Mukhi, C. Schubert and A. Sengupta for the
discussions at different stages of this work.

\refout
\end